\documentclass[aps,prb,twocolumn,groupedaddress,10pt,longbibliography]{revtex4-2}
\usepackage{graphicx}
\usepackage{dcolumn}
\usepackage{bm}
\usepackage[english]{babel}
\usepackage{amsmath}
\usepackage[utf8]{inputenc}
\usepackage[T1]{fontenc}
\usepackage{etoolbox}
\usepackage{float}
\usepackage{placeins}

\makeatletter
\def\@email#1#2{%  
 \endgroup
 \patchcmd{\titleblock@produce}
  {\frontmatter@RRAPformat}
  {\frontmatter@RRAPformat{\produce@RRAP{*#1\href{mailto:#2}{#2}}}\frontmatter@RRAPformat}
  {}{}
}%
\makeatother

\begin{document}
\preprint{APS/123-QED}

\title{Measuring the magnetic anisotropy of the spin Hall effect and spin diffusion length in nickel and permalloy via electrical spin injection}

\author{Eoin Dolan$^{1,2}$, Jone Mencos$^{1,2}$, Williams Savero Torres$^{4}$, Maxen Cosset-Chéneau$^{5}$, Jean-Philippe Attané$^{4}$, Laurent Vila$^{4}$,  Luis E. Hueso$^{1,3}$, Fèlix Casanova$^{1,3*}$}
\affiliation{$^1$CIC nanoGUNE BRTA, 20018 Donostia-San Sebastian, Basque Country, Spain}
\affiliation{$^2$Departamento de Polímeros y Materiales Avanzados: Física, Química y Tecnología, University of the Basque Country (UPV/EHU), 20018 Donostia-San Sebastian, Basque Country, Spain}
\affiliation{$^3$IKERBASQUE, Basque Foundation for Science, 48009 Bilbao, Basque Country, Spain}
\affiliation{$^4$University Grenoble Alpes, CEA, CNRS, Grenoble INP, SPINTEC, 38000 Grenoble, France}
\affiliation{$^5$Zernike Institute for Advanced Materials, University of Groningen, 9747 AG Groningen, The Netherlands}

\email[Author to whom correspondence should be addressed: ]{f.casanova@nanogune.eu}

\begin{abstract}
The spin Hall effect in ferromagnets is of great interest in the field of spintronics, and while the effect has been quantified in many materials, the dependence of the spin Hall angle on the relative orientation of spin polarization and the magnetization is less well studied. Of equal importance for the purpose of spin-charge interconversion in ferromagnets is the spin relaxation length, which is predicted to be highly anisotropic with respect to magnetization. Using a modified lateral spin valve geometry with a copper channel and permalloy spin injector, we measure the dependence of the spin Hall angle and spin relaxation length on magnetization orientation in permalloy and nickel, using two distinct device geometries. This allows us to disentangle the contributions of the spin relaxation length and spin Hall angle to the measured spin-charge interconversion voltage output. Our results indicate a large anisotropy in both the spin relaxation length and spin Hall angle in both permalloy and nickel, in agreement with theoretical calculations. The quantities change in opposite directions, with the spin relaxation length rising as the magnetization is moved parallel to the spin polarization and the spin Hall angle falling, leading to a near total cancellation of the spin-charge interconversion output.
\end{abstract}

\maketitle
\section{Introduction}
The spin Hall effect (SHE) allows for the conversion of charge current into spin current and vice-versa, and is therefore of fundamental importance in the field of spintronics \cite{sinova_spin_2015}. In typical spintronic devices, non-magnetic heavy metals are used where components with strong SHE are required, with ferromagnets (FM) being reserved for memory elements such as in MRAM \cite{Kim2024-mw, Shao2021-te}, and spin-based logic \cite{manipatruni_scalable_2019, Incorvia2024-ja, vaz_voltage-based_2024}. However, FMs can display SHE \cite{qin_nonlocal_2017, iihama_spin-transfer_2018, gibbons_reorientable_2018, omori_relation_2019, cosset-cheneau_electrical_2022}, with spin-charge interconversion (SCI) efficiencies comparable to heavy metals \cite{cosset-cheneau_electrical_2022}, and with the potential advantage of using their magnetization as an additional degree of freedom for control for SCI \cite{miura_first-principles_2021, cosset-cheneau_electrical_2022}.

The spin Hall conductivity is expected to be anisotropic in FMs due to lowering of symmetry by the magnetization \cite{Salemi2022, Seemann2015-xj}. This is sometimes referred to as the spin anomalous Hall effect (SAHE), although we discuss the effect in terms of anisotropy in the SHE \cite{Taniguchi2015-eg, amin_intrinsic_2019}. Where the SAHE is mentioned we are referring to literature sources which use this term \cite{Salemi2022, Zheng2024-tn}. A spin current in a FM is expected to produce a charge current which depends on the relative orientation of the spin with the magnetization, which defines the direction of spin polarization in a FM at equilibrium. It has been speculated that the SHE and AHE can be related to one another by the FM polarization \cite{tsukahara_self-induced_2014}, but in practice, the relationship between the SHE and AHE in a FM is quite complex \cite{Salemi2022}, and thus the SHE should be considered in terms of the spin Hall conductivity \cite{omori_relation_2019, Salemi2022, Seemann2015-xj}.

The anisotropy of the SHE has been studied in several different FMs, such as CoFeB \cite{iihama_spin-transfer_2018, PhysRevB.99.104414}, FePt \cite{seki_large_2019}, CoFe \cite{koike_composition_2020}, NiFe (Permalloy, Py) \cite{das_efficient_2018, qin_nonlocal_2017, das_spin_2017, Zhu2020-ao, Seki2021-kq}, Co \cite{tian_manipulation_2016}, CoTb \cite{Yagmur2020-ul}, NiCu \cite{cosset-cheneau_electrical_2022, Varotto2020-qa, Cheng2022-wq}, NiPd \cite{cosset-cheneau_electrical_2022}, and Fe \cite{Zhu2023} using a variety of measurement techniques. Results are mixed with some studies showing the SHE to be isotropic \cite{cosset-cheneau_electrical_2022, tian_manipulation_2016, Varotto2020-qa, Cheng2022-wq} and others showing it to be anisotropic \cite{iihama_spin-transfer_2018, PhysRevB.99.104414, seki_large_2019, koike_composition_2020, das_efficient_2018, das_spin_2017, Zhu2020-ao, Yagmur2020-ul, Zhu2023}--whether this is due to experimental differences or material dependent is yet to be established. Theoretical studies of Ni predict a large difference in the SHE with magnetization \cite{amin_intrinsic_2019, qu_magnetization_2020, Salemi2022}, which should be easily resolvable in our experiment. In Py, prior experimental work \cite{das_spin_2017, das_efficient_2018} has shown an appreciable anisotropy. However prior works on this topic do not typically consider a dynamic spin relaxation length (\(\lambda_\mathrm{s}\)) which is also expected to vary greatly with magnetization orientation in both Py and Ni \cite{petitjean_unified_2012}, meaning the effects of \(\lambda_\mathrm{s}\) and the spin Hall angle (\(\theta_{\text{SHE}}\)) are entangled. 

In this work, we study the SHE in Py and Ni, using lateral spin valves (LSVs) in a non-local configuration to produce a pure spin current. Note that we measure the inverse SHE, but the direct SHE can be obtained by swapping the current and voltage terminals, which follows Onsager reciprocity \cite{Buttiker1988-ub, Kimura2007-bu}. For the sake of simplicity, we will refer to both as SHE. This method of spin injection/detection provides for a clear and unambiguous analysis of the results \cite{qin_nonlocal_2017}. This geometry allows us to clearly separate the spin injecting and detecting FMs, reducing the chances of spurious effects from spin injection, and eliminating the chance of exchange interaction between the FMs. We employ a geometry which allows us to measure the anisotropy of $\theta_\mathrm{SHE}$ with magnetization, by independently controlling the magnetization of the spin injecting and detecting FMs, while simultaneously accounting for the anisotropy of $\lambda_\mathrm{s}$ in the FM. In this way, we observe a large anisotropy of both $\theta_\mathrm{SHE}$ and $\lambda_\mathrm{s}$, depending on the relative orientation of the spin polarization of the FM magnetization. Crucially, with the spin polarized perpendicular (parallel) to the magnetization, we see a minimum (maximum) in $\lambda_\mathrm{s}$ and a maximum (minimum) in $\theta_\mathrm{SHE}$, with the product of the two, and therefore the measured voltage due to SHE, being approximately constant.

\begin{figure}[h!]
    \centering
    \includegraphics[width=0.48\textwidth]{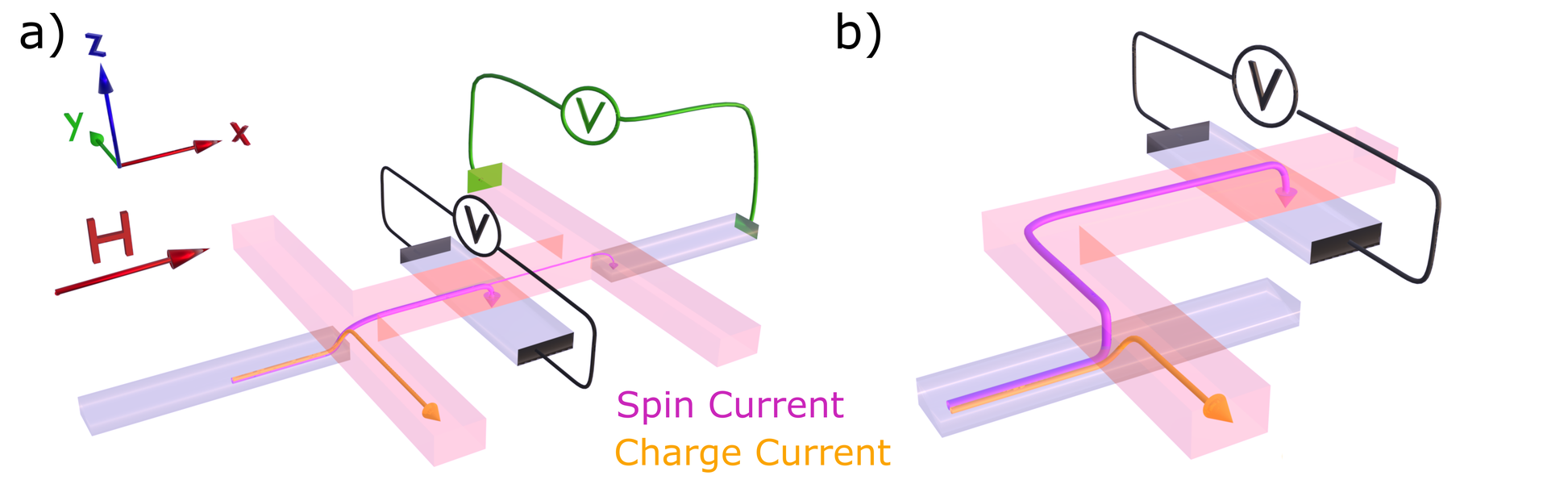}
    \caption{\label{fig:device_schematic}
    Device Schematic. Two geometries were used in this work to measure the SHE.
    \textbf{(a)} H-shaped device, consisting of two Py elements (grey) along $\hat{x}$, referred to as the injector (left) and detector (right), with a third perpendicular element (absorber, centre) along $\hat{y}$ made of either Py or Ni, all connected by the Cu channel (orange). Two voltage measurements are indicated: in black is the SHE configuration (corresponding to results in Fig. 3), and in green the spin absorption configuration (corresponding to results in Fig. 2).
    \textbf{(b)} L-shaped device, again showing a Py injector (along $\hat{x}$), and a Py or Ni absorber (along $\hat{y}$) with an associated spin Hall voltage (corresponding to results in Fig. 4).}
\end{figure}

\section{Device}

We study Py and Ni, prototypical FM examples, which both form a high-quality transparent interface with the Cu we use for our spin channel. Our measurements indicate that the interface resistance between the two materials is in fact limited by the Sharvin resistance, which forms a key part of our analysis \cite{BRATAAS2006, cosset-cheneau_measurement_2021}. Py is used as the spin injecting material in all devices, since Ni cannot be used to inject or detect spin current in LSVs, with it having zero current spin polarization \cite{jedema_spin_2003, PhysRevLett.83.1427}.

Our FMs are deposited by electron-beam deposition onto a clean Si/SiO$_x$ substrate using standard lithographic techniques. All FM elements are 30~nm thick and 140~nm wide, with lengths of $\bm{\approx 2\,\mathrm{\mu m}}$. The spin injector and detector have different switching fields due to engineered shape anisotropies (the addition of a nucleation pad far from the Cu/Py junction favors magnetic domain wall formation, lowering the switching field). After depositing the FM elements, the lithography defining the Cu channel, either H-shaped, Fig.\,\ref{fig:device_schematic}a, or L-shaped, Fig.\,\ref{fig:device_schematic}b, was performed. Before Cu deposition, the interfaces of the FMs were cleaned using Ar-ion milling to ensure a transparent and Ohmic interface, and a Cu channel (140~nm wide and 90~nm thick) was deposited by thermal evaporation in ultra-high vacuum. All lateral dimensions were confirmed by scanning electron microscopy (SEM), and thicknesses were confirmed by X-ray diffraction (XRD). The SHE element of the devices (henceforth referred to as the absorber), indicated with the black voltage probe in Fig.~\ref{fig:device_schematic}(a,b), is either Py or Ni, with both materials studied in the same geometries.

First, we characterized our material properties, including bulk resistivities ($\rho$), which were obtained using four-point electrical measurements combined with device dimensions from SEM imaging. We calculated the interface resistances between Cu/Py and Cu/Ni \cite{Pham2021-fh}, although this was found to be smaller than our experimental resolution and in fact the intrinsic Sharvin resistance is likely the dominant interface term (Note S3 \cite{supinfo}). The saturation magnetic fields for Py and Ni were found via anisotropic magnetoresistance (AMR) measurements (Note S4 \cite{supinfo}). We also measured, using a separate reference device with a conventional LSV geometry, the spin relaxation length for a spin current polarization oriented parallel to the magnetization ($\lambda_\mathrm{s}^{\parallel}$) in Py, as well as the spin polarization of Py ($\alpha_\mathrm{Py}$), which are shown in Table I (see Notes S1 and S2 for details \cite{supinfo}). These were extracted using 3D FEM simulations with GetDP \cite{getdp} and Gmsh \cite{Geuzaine2009-uf}, of LSVs with different channel lengths, together with data from an LSV with an absorbing Py element \cite{sagasta_spin_2017, zahnd_comparison_2016}. This is particularly important because Py will be used to inject spin in both the H-shaped and L-shaped devices, with the injector always oriented parallel to the magnetic field, so we can use the values derived from the reference device in the Py elements aligned with the magnetic field.

\vspace{-1em}
\begin{table}[!htb]
    \centering
    \caption{Key material properties extracted from reference measurements and simulations at 10\,K.}
    \renewcommand{\arraystretch}{1.2} 
    \setlength{\tabcolsep}{5pt} 
    \begin{tabular}{l c c c}
        \hline\hline
        \textbf{Material} & $\lambda_{\text{s}}^\mathrm{\parallel}$ (nm) & \textbf{$\rho$} ($\Omega \cdot \text{nm}$) & \textbf{$\alpha$} \\ 
        \hline
        Py & $2.6 \pm 0.2$ & $313 \pm 5$ & $0.33 \pm 0.02$ \\ 
        Cu & $1350 \pm 50$ & $19 \pm 2$ & $0$ \\ 
        \hline\hline
    \end{tabular}
    \label{tab:material_properties}
    \vspace{-2em} 
\end{table}
\vspace{-1em} 

\begin{figure*}[!htp]
    \centering
    \includegraphics[width=1\textwidth]{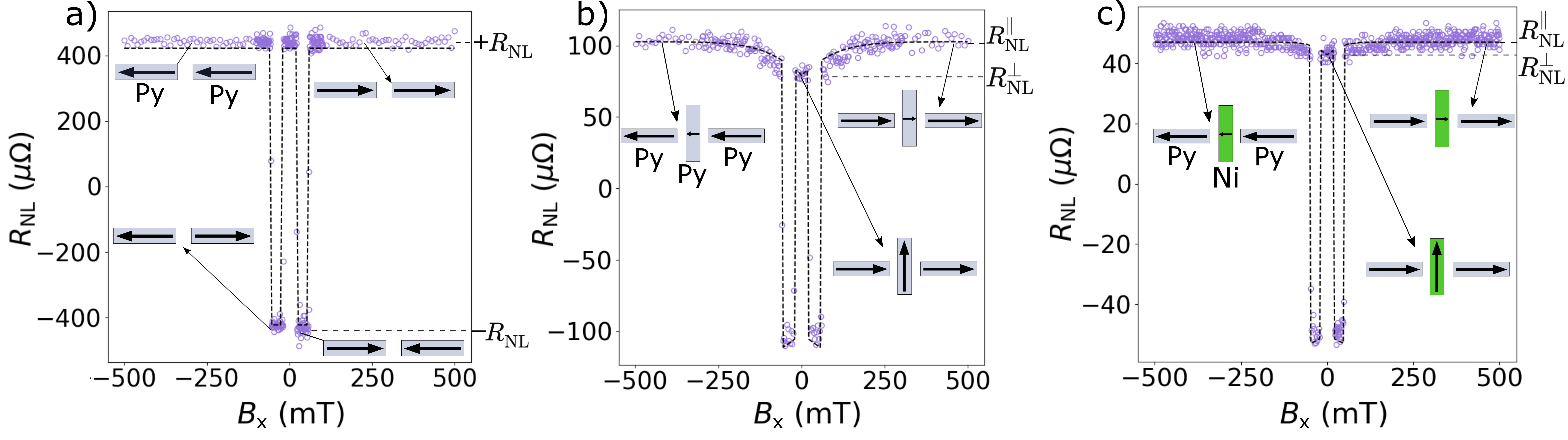}
    \caption{\label{fig:absorption} 
    Spin absorption anisotropy measurements using the H-shaped device, with a channel width of 130\,nm and injector/detector distance of 500\,nm
    \textbf{(a)} A reference LSV measurement, made in a device identical to that in Fig.~\ref{fig:device_schematic}(a), but without the central FM absorber. The relative orientations of the FM injector and detector magnetizations are indicated, corresponding to two measured non-local resistances, $\pm R_{\text{NL}}$, for parallel/antiparallel magnetizations.
    \textbf{(b, c)} LSV measurements with an absorbing Py \textbf{(b)} and Ni \textbf{(c)} element placed between the injector and detector. The signal at high field is indicated as $R_{\text{NL}}^{\parallel}$ (spins polarized parallel to the absorber magnetization), and at zero field as $R_{\text{NL}}^{\perp}$ (spins polarized perpendicular to the absorber magnetization). The black dashed line represents a fit from 3D FEM simulation.}

\end{figure*}

\section{Spin Absorption Anisotropy}
The aim of this section is to quantify the spin relaxation anisotropy by measuring the spin absorption anisotropy. To do this we employ an LSV configuration in the H-shaped device both with (Fig.~\ref{fig:device_schematic}(a), green circuit) and without an absorber. This is distinct from the analysis we do to extract $\lambda_{\text{s}}^{\parallel}$ in Py (Table I) as shown in Note S2 \cite{supinfo}, because now the absorber is no longer magnetized parallel/anti-parallel to the injector and detector, except at very high field. The absorber will still absorb spin, reducing the spin accumulation at the detector, but this absorption now depends on the applied magnetic field, which rotates the absorber magnetization from being perpendicular to the injector/detector at zero field to being parallel at high field. The absorption has been shown to be much more efficient in the perpendicular orientation \cite{petitjean_unified_2012}, because the spins entering the absorber precess and dephase in the exchange field of the absorber. This results in a smaller signal at low field, where absorption is more efficient and little spin reaches the detector.

The results, showing the non-local resistance $R_\mathrm{NL}=V_\mathrm{NL}/I_\mathrm{c}$, where $V_\mathrm{NL}$ and $I_\mathrm{c}$ are the measured non-local voltages and applied charge currents (green circuit, Fig.~\ref{fig:device_schematic}a), are plotted in Fig.~\ref{fig:absorption}b,c. They demonstrate strong spin current absorption in both Py and Ni compared to the reference device (Fig.~\ref{fig:absorption}a). At zero applied magnetic field (perpendicular configuration), Py absorbs $84 \pm 1\%$ and Ni absorbs $92 \pm 1\%$ of the spin current. In the parallel configuration, the absorption decreases to $78 \pm 1\%$ for Py and $89 \pm 1\%$ for Ni.

In the 3D FEM model, this is simulated by a changing value of the spin relaxation length, with $\lambda_\mathrm{s}^\mathrm{}$ being much shorter for spins oriented perpendicular to the magnetization. This value is denoted $\lambda_{\text{s}}^{\perp}$, where we expect $\lambda_{\text{s}}^{\parallel} > \lambda_{\text{s}}^{\perp}$. $\lambda_{\text{s}}^{\parallel}$ in Ni is extracted from the H-shaped device. Given that we know the value of $\lambda_{\text{s}}^{\parallel}$ in Py as well as $\alpha_\mathrm{Py}$, we can use these in the 3D FEM simulation for the Py injector and detector, and vary $\lambda_{\text{s}}^{\parallel}$ in Ni until we match the high field experimental $R_\mathrm{NL}$ signal in Fig. \ref{fig:absorption}c.

To find the value of $\lambda_\mathrm{s}^{\perp}$ we employ the same approach, but now focusing on $R_\mathrm{NL}$ at low field. We can keep the same value of $\lambda_\mathrm{s}^{\parallel}$ for the injector and detector as we found from the reference device. Indeed, at high field we can also use these values for the absorber, which gives a good fit to the data at high field (Fig. \ref{fig:absorption}b), although it deviates slightly in the H-shaped reference device (no absorber, Fig. \ref{fig:absorption}a). To fit the signal at low field, we vary $\lambda_\mathrm{s}^{\perp}$ until we match the experimental $R_\mathrm{NL}$. For the fitting curves shown in Fig. \ref{fig:absorption}b and c, for Py and Ni respectively, intermediate values are fitted simply by interpolating between $\lambda_\mathrm{s}^{\parallel}$ and $\lambda_\mathrm{s}^{\perp}$, following the known shape of the magnetization derived from AMR data (Note S4 \cite{supinfo}). 

The values for $\lambda_\mathrm{s}^{\parallel}$ and $\lambda_\mathrm{s}^{\perp}$ extracted in this way are summarized in Table~\ref{tab:Absorption}, noting that the same value of $\lambda_\mathrm{s}^{\parallel}$ is used in the reference device and the H-shaped device for Py. The absorption clearly depends on the magnetization of the FM, although this effect is much stronger in Py (Fig.~\ref{fig:absorption}b), it is also visible in Ni (Fig.~\ref{fig:absorption}c).

\begin{table}[H]
    \centering
    \caption{Spin relaxation lengths, spin Hall angles (in units of $\hbar/e$), resistivities, and $\theta \lambda$ products at 10\,K.}
    \footnotesize
    \renewcommand{\arraystretch}{1.2}
    \setlength{\tabcolsep}{4pt}
    \resizebox{\columnwidth}{!}{%
    \begin{tabular}{l c c c c c c c}
        \hline\hline
        \textbf{} 
        & $\rho$ ($\Omega$\,nm) 
        & $\lambda_{\text{s}}^{\parallel}$ (nm) 
        & $\lambda_{\text{s}}^{\perp}$ (nm) 
        & $\theta_{\text{SHE}}^{\parallel}$ (\%) 
        & $\theta_{\text{SHE}}^{\perp}$ (\%) 
        & $\theta^{\parallel} \lambda^{\parallel}$ (nm)
        & $\theta^{\perp} \lambda^{\perp}$ (nm)\\
        \hline
        Py & $313 \pm 5$ & $2.6 \pm 0.2$ & $0.9 \pm 0.2$ & $1.1 \pm 0.2$ & $4.2 \pm 0.5$ & $0.029 \pm 0.007$ & $0.038 \pm 0.013$ \\
        Ni & $115 \pm 2$ & $3.1 \pm 0.2$ & $2.0 \pm 0.2$ & $2.3 \pm 0.3$ & $3.5 \pm 0.4$ & $0.071 \pm 0.014$ & $0.07 \pm 0.015$ \\
        \hline\hline
    \end{tabular}
    }
    \label{tab:Absorption}
    \vspace{-1em}
\end{table}

Although the difference between $R_\mathrm{NL}^\mathrm{\parallel}$ and $R_\mathrm{NL}^\mathrm{\perp}$ is rather small in both cases, the values for $\lambda_\mathrm{s}^{\parallel}$ and $\lambda_\mathrm{s}^{\perp}$ extracted from our simulations differ much more. This is due to the presence of the Sharvin resistance at the Cu/FM interfaces, which is relevant for both $\lambda_\mathrm{s}^{\parallel}$ and $\lambda_\mathrm{s}^{\perp}$, although much more so in the latter case. The Sharvin resistance provides a maximum conductance value of the interface. For a spin potential in the Cu channel, this acts as a resistance in series with the spin resistance of the Py injector/detector, and at short spin relaxation length in the FM absorber (below 5 nm) the Sharvin resistance contributes appreciably to the absorption. For analysis in this work, we use the value of $ 1.2 \times 10^{15} \, \text{f} \Omega^{-1} \text{m}^{-2} $ for all interfaces, based on $ab\,\,initio$ studies of Cu/3d-FM interfaces \cite{BRATAAS2006, cosset-cheneau_measurement_2021}, although a range of values can be found in the literature \cite{Bauer2003-fs, BRATAAS2006, petitjean_unified_2012, cosset-cheneau_measurement_2021}. This has a significant effect on the absolute values for both $\lambda_\mathrm{s}$ and $\theta_\mathrm{SHE}$ we calculate. More detailed discussion can be found in Note S2 \cite{supinfo}. 
In this experiment, we observe a very small difference in values for $\lambda_\mathrm{s}^{\parallel}$ between Py and Ni (Table II). Theoretical predictions expect a much larger value of $\lambda_\mathrm{s}^{\parallel}$ in Ni due to its weaker exchange splitting (as high as 20\,nm), as well as a somewhat larger value of $\lambda_\mathrm{s}^{\parallel}$ in Py (5.5\,nm) \cite{petitjean_unified_2012}, although these predictions are all for monocrystalline materials, and in our case we have polycrystalline materials, which along with various experimental inevitabilities can be expected to lower $\lambda_\mathrm{s}$ compared to the theoretical ideal. For both materials, $\lambda_\mathrm{s}^{\perp}$ is expected to be very small (< 1\,nm), due to rapid precession and dephasing of the spin current over a length scale on the order of a few interatomic distances \cite{carva_spin-mixing_2007, petitjean_unified_2012}. Indeed we do observe a reduction of $\lambda_\mathrm{s}^{\perp}$ compared to $\lambda_\mathrm{s}^{\parallel}$ in both materials, although given the very small values involved for $\lambda_\mathrm{s}^{\perp}$ it is difficult to assign an accurate value experimentally, particularly given the dependence on the interface resistance which cannot be precisely determined (Note S2 \cite{supinfo}).

\section{Spin Hall Effect anisotropy}

\begin{figure}[h!]
    \centering
    \includegraphics[width=0.48\textwidth]{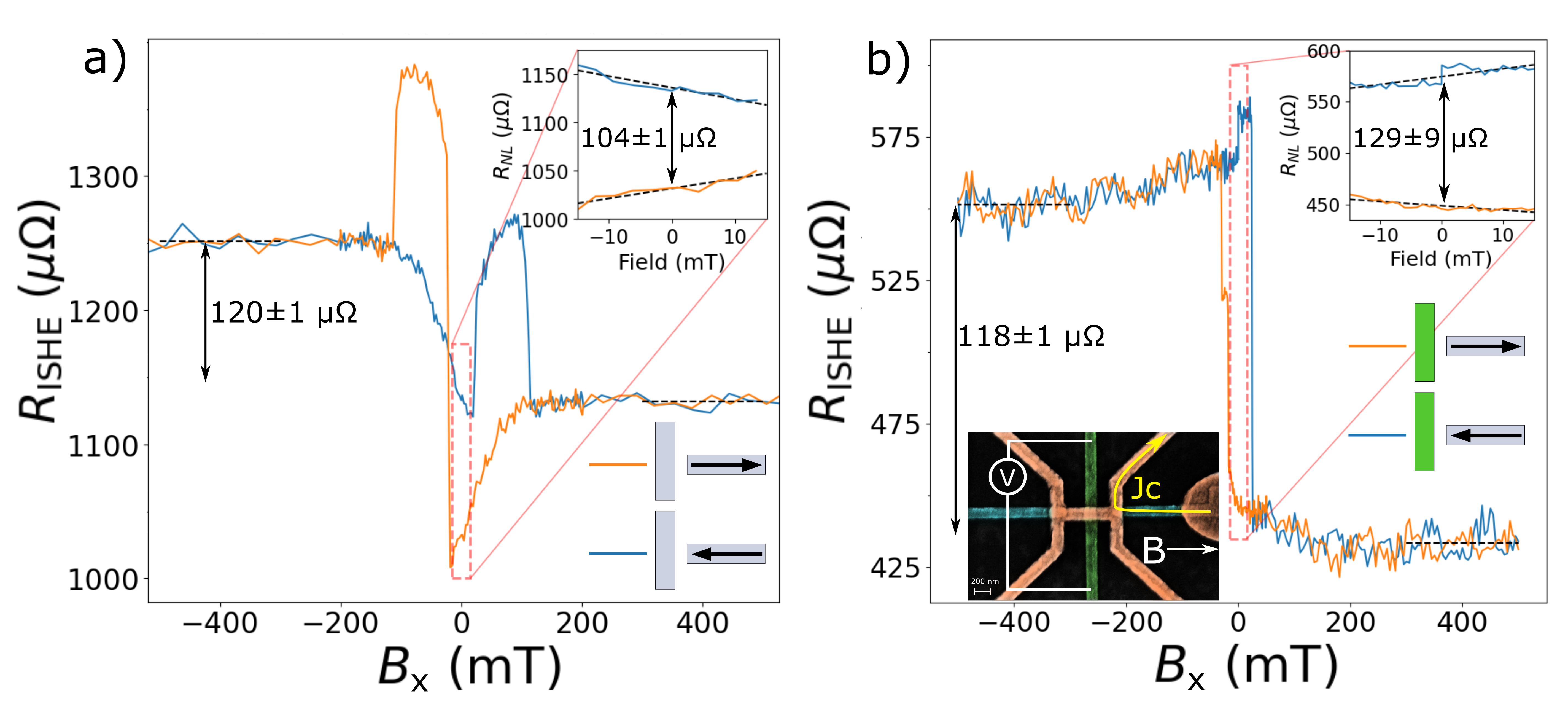}
    \caption{\label{fig:LASH} SHE anisotropy measurement in \textbf{(a)} Py and \textbf{(b)} Ni, using the H-shaped device. The inset shows a false-colour SEM image of the device used, indicating the applied field $B_\mathrm{x}$, applied charge current $J_\mathrm{c}$ and the measured voltage. The orange and blue curves correspond to the two initial magnetization states of the injector, with the top right insets showing zooms at zero field. $2\Delta R_{\mathrm{ISHE}}$ at zero field and high field is annotated.}
\end{figure}

In the same H-shaped device used for measuring absorption anisotropy, we can also measure the SHE (Fig.~\ref{fig:device_schematic}a, black circuit). In this configuration, the injector magnetization, and therefore the spin polarization, is fixed along $\hat{x}$. Similarly, the direction of the spin current ($J_\mathrm{s}$) flows almost entirely along $\hat{z}$, from the Cu channel into the FM absorber. The spin Hall current is therefore generated along $\hat{y}$, corresponding to the measured voltage. The magnitude of the effect, $2\Delta R_{\mathrm{ISHE}}$, corresponding to the difference in the signal, $R_\mathrm{ISHE}=V_\mathrm{ISHE}/I_\mathrm{c}$, where $V_\mathrm{ISHE}$ is the voltage due to the SHE, for the injector magnetized along $+\hat{x}$ and $-\hat{x}$ is shown in Fig.~\ref{fig:LASH} at high field and zero field. 

To perform the measurement, the injector and absorber are first magnetized along 
$\pm \hat{x}$ and $\pm \hat{y}$, respectively, by the application of a magnetic field along $\hat{y}$. Then the field is swept along $\hat{x}$, from 0 to $\pm 500~\mathrm{mT}$, while measuring the voltage. This process is repeated for positive and negative field sweeps, saturating the injector and detector before each sweep. In practice, and in agreement with theory, the initial orientation of the absorber ($+\hat{y}$ or $-\hat{y}$) does not affect the result, since any anisotropy with this symmetry depends on the direction but not the sign of the absorber magnetization \cite{amin_intrinsic_2019}. $R_{\mathrm{ISHE}}$ therefore does not depend on whether the absorber is initially magnetized along $+\hat{y}$ or $-\hat{y}$, but only on $\sin^2\beta$, where $\beta$ is the angle between the spin polarization and the magnetization.

Ideally, if both $\lambda_\mathrm{s}$ and $\theta_\mathrm{SHE}$ were isotropic with respect to magnetization, the SHE measurement should produce a square hysteresis loop, with switching corresponding to the switching of the injector. This is because the injector switching changes the sign of the spin polarization, therefore changing the sign of $V_\mathrm{ISHE}$, and thus $R_\mathrm{ISHE}$. In the realistic case, we would expect to see an increase in the signal at high field as the value of $\lambda_\mathrm{s}$ increases (see Fig.~\ref{fig:absorption}), as well as the contribution from any change in $\theta_{\mathrm{SHE}}$ with respect to zero-field. In the case of Ni, this is indeed the case; except for jumps at low field, the signal matches what we would expect in a system where $\theta_{\mathrm{SHE}} \times \lambda_\mathrm{s}$ is approximately constant over the range of field measured.

In contrast, the data from the Py measurement is clearly quite different. There are several possible interpretations here, with artefacts possible from the planar Hall effect and anomalous Hall effect (AHE). However, these contributions would also affect Ni. Therefore, the most likely cause is the detection of spin accumulation at the interface of the FM absorber and Cu, which is measurable in Py but not in Ni. The etching of the absorber to give a clean interface with the Cu induces asymmetric domain trapping centres on both sides of the Cu channel. The magnetic texture is therefore not the same at these two sides, leading to a measurement of a non-local signal unrelated to spin-charge interconversion effects \cite{cosset-cheneau_electrical_2022, Savero_Torres2016-qr}. A spin accumulation measurement in Py is significantly larger than the SHE -- when measuring the spin accumulation directly in the same device, we find a signal with the same symmetry as the background, but much larger (see Note S5 \cite{supinfo}). Detecting even a small part of this signal would be comparable to the SHE, which matches what we see here.

These spin accumulation contributions are symmetric at high magnetic field, allowing us to easily disentangle them from the anti-symmetric SHE signal. At low field the spin accumulation does not contribute, meaning we directly measure just the SHE (Note S5 \cite{supinfo}). The value of $2 \Delta R_{\mathrm{ISHE}}$ is obtained by fitting a flat line to the $R_{\mathrm{ISHE}}$ values at high positive and negative field after saturation of the absorber, and subtracting the values. Similarly, at low field, $2 \Delta R_{\mathrm{ISHE}}$ is obtained by fitting a line to the low-field values of $R_{\mathrm{ISHE}}$, thus determining the zero-field non-local resistance. Regardless of the behaviour at intermediate fields, complicated by the presence of spin accumulation in the case of Py, the values at low and high fields are sufficient to measure the anisotropy of $\theta_{\mathrm{SHE}}$. In this H-shaped device we measure $2 \Delta R_{\mathrm{ISHE}}$ at zero field and high field of $104 \pm 1\,\mathrm{\mu\Omega}$ and $120 \pm 1\,\mathrm{\mu\Omega}$ respectively for Py (Fig. \ref{fig:LASH}a), and $129 \pm 9\,\mathrm{\mu\Omega}$ and $118 \pm 1\,\mathrm{\mu\Omega}$ for Ni (Fig. \ref{fig:LASH}b)

In this device geometry, the FM elements can be placed very close together, allowing for a large signal to be measured (Fig.~\ref{fig:LASH}). However, this proximity can lead to unwanted artefacts due to an appreciable fraction of the charge current reaching the absorber, although any artefacts from the AHE or magnetoresistance would be symmetric with magnetization and thus easily removed. In addition, the current is injected through the tip of the injector, and, although the magnetization is, in principle, still oriented along $\hat{x}$ at the tip, there is the potential for some non-collinear component, especially at low fields - although such effects are not seen in the reference device, Fig. \ref{fig:absorption}a.

\begin{figure}[h!]
    \centering
    \includegraphics[width=0.48\textwidth]{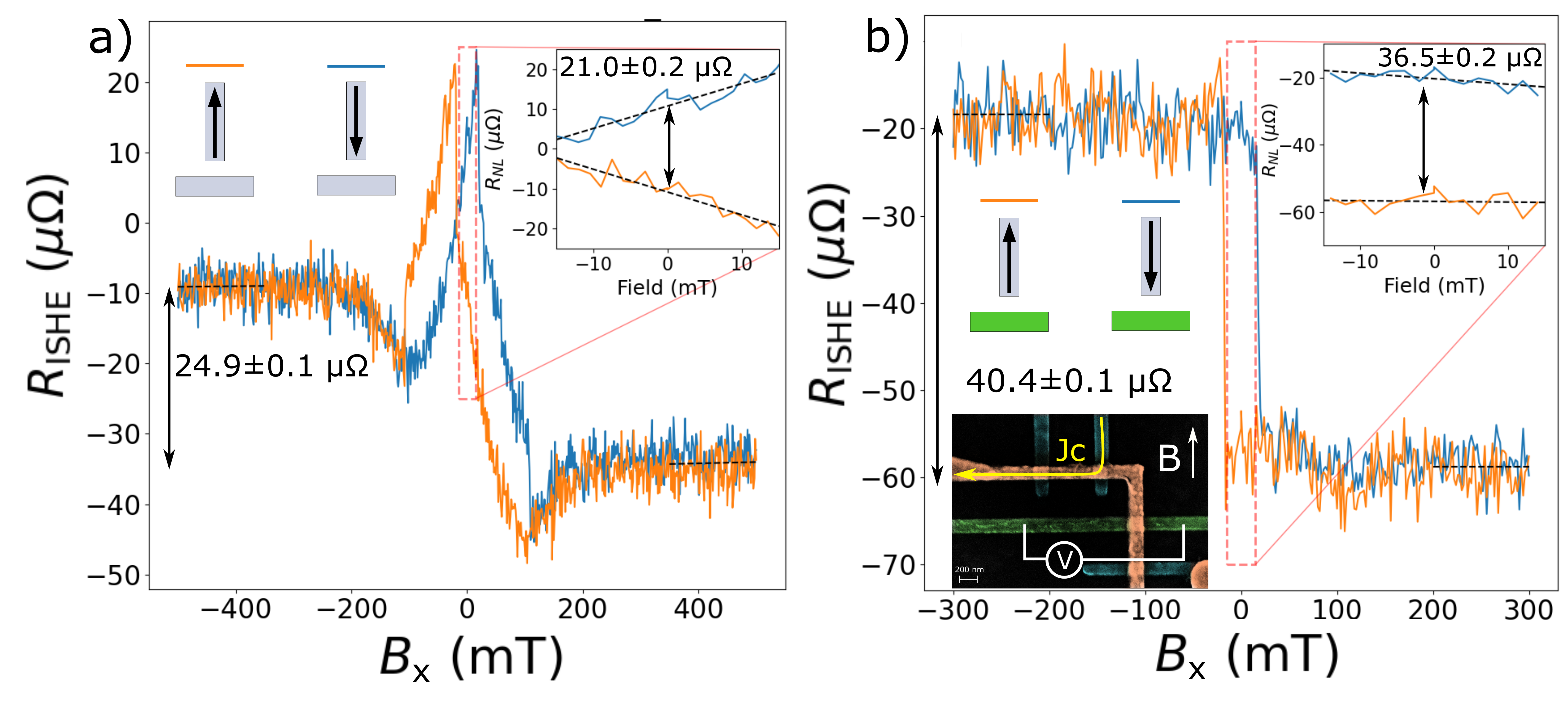}
    \caption{\label{fig:Corner} SHE anisotropy measurement in \textbf{(a)} Py and \textbf{(b)} Ni using the L-shaped device. The inset shows a false-colour SEM image of the devices used, indicating the applied field $B_\mathrm{x}$, applied charge current $J_\mathrm{c}$ and the measured voltage. The orange and blue curves correspond to the two initial magnetization states of the injector, with the top right insets showing zooms at zero field. $2\Delta R_{\mathrm{ISHE}}$ at zero field and high field is annotated.}
\end{figure}

To overcome these potential issues, we carry out the same measurement using a different geometry, the L-shaped device shown in Fig.~\ref{fig:device_schematic}(b). In this case, the FMs cannot be brought so close together, meaning a lower signal, but it does allow for current injection through the body of the injector rather than the tip, as well as reduced charge current at the absorber, both of which lead to reduced potential spurious effects. Although this second geometry gives a cleaner signal, it is clearly not perfect in the case of Py, where a spin accumulation signal is again seen (Fig. 4a). For Ni, we get a square signal (Fig. 4b), as expected in the case of isotropic SHE in this configuration. Despite their different shapes, the two geometries share the same fundamental principle, with a perpendicularly oriented injector/absorber pair and a Cu channel allowing for non-local spin injection. In this L-shaped device we measure $2 \Delta R_{\mathrm{ISHE}}$ at zero field and high field of $21.0 \pm 0.2\,\mathrm{\mu\Omega}$ and $24.9 \pm 0.1\,\mathrm{\mu\Omega}$ respectively for Py (Fig. \ref{fig:Corner}a), and $36.5 \pm 0.2\,\mathrm{\mu\Omega}$ and $40.4 \pm 0.1\,\mathrm{\mu\Omega}$ for Ni (Fig. \ref{fig:Corner}b).

\begin{figure}[h!]
  \centering
  \includegraphics[width=0.48\textwidth]{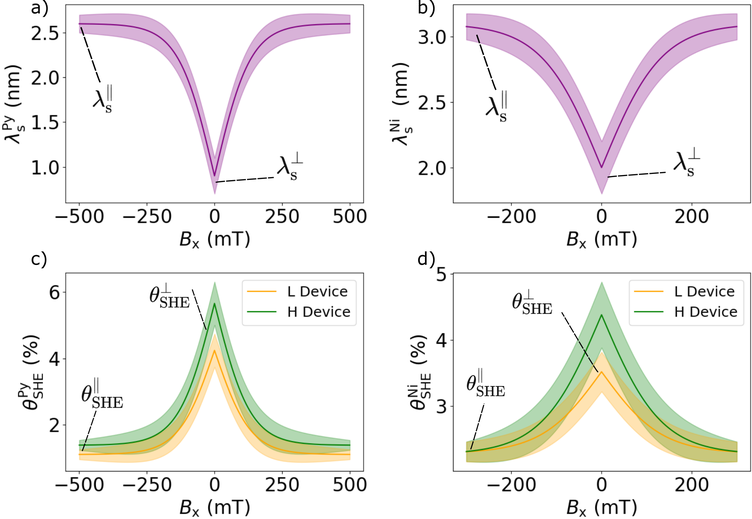}
  \caption{\label{fig:Results} 
  Magnetic field dependence of $\lambda_\mathrm{s}$ in Py (a) and Ni (b) and $\theta_{\mathrm{SHE}}$ in Py (c) and Ni (d). The values of $\lambda_\mathrm{s}$ are obtained by fitting the data in Figs.~2b and c (H-shaped device), while $\theta_{\mathrm{SHE}}$ is extracted from fitting the data in Figs.~3a and b (H-shaped device) and Figs.~4a and b (L-shaped device), in both cases using $\lambda_\mathrm{s}$ as an input}
\end{figure}

We calculated the value of $\theta_{\mathrm{SHE}}$ for Py (Fig.~5c) and Ni (Fig.~5d) in both device geometries (Fig.~1a,b). The two return very similar results for $\theta_{\mathrm{SHE}}$. This calculation is again performed via a 3D FEM simulation of each device, using the known value of $\lambda_\mathrm{s}$ at each field as an input parameter, and treating $\theta_{\mathrm{SHE}}$ as a fitting parameter. The value of $\theta_{\mathrm{SHE}}$ is iterated until the simulated $R_{\mathrm{ISHE}}$ at both low and high field matches the experimental measurements, with values at intermediate fields interpolated between these two values, given the known shape of the magnetization pulling extracted from AMR measurements of the FM absorber (Note S4 \cite{supinfo}). The values of $\theta_{\mathrm{SHE}}$ shown in Table II are from the L-shaped device.

We observe a large anisotropy in $\theta_\mathrm{SHE}$ in both materials, although it is much larger in Py than in Ni. Interestingly, despite this very large anisotropy, the value of $R_\mathrm{ISHE}$ changes only slightly as a function of magnetization direction. This is due to the compensation of $\theta_\mathrm{SHE}$ and $\lambda_\mathrm{s}$, where $R_\mathrm{ISHE} \propto \theta_\mathrm{SHE} \times \lambda_\mathrm{s}$. In these materials, we appear to get an almost complete cancellation between the two quantities (i.e., $R_\mathrm{ISHE}^\parallel \approx R_\mathrm{ISHE}^\perp$) with $\lambda_\mathrm{s}$ being reduced greatly for spins perpendicular to the magnetization as expected, but this being compensated by a much larger value of $\theta_\mathrm{SHE}$. Equivalently, we can say that the product, $\theta_\mathrm{SHE} \times\lambda_\mathrm{s}$, which is often used to express the overall efficiency of the SHE in a system is approximately constant for all values of magnetization, i.e. $\theta_\mathrm{SHE}^\parallel \times \lambda_\mathrm{s}^\parallel$ $\approx$ $\theta_\mathrm{SHE}^\perp \times \lambda_\mathrm{s}^\perp$, for both Py and Ni. The same result has been observed in the literature for NiCu and NiPd \cite{cosset-cheneau_electrical_2022}.

Comparing this to literature measurements, we see quite different results, with an experiment using magnon transport in Py observing $\Delta R_\mathrm{ISHE}$ changing by a factor of two between the parallel and perpendicular spin states \cite{das_spin_2017}. For Ni no such experiments are available to compare with our results, however as a pure element it is comparatively better theoretically characterized than Py. In Ref. \cite{amin_intrinsic_2019}, the anisotropy of the SHE is decomposed into a magnetization independent SHE along with a SAHE, which arises when the spin polarization is aligned with the magnetization, which is equivalent to considering an anisotropy of the SHE with the symmetry of the SAHE as we do here. A similar calculation is carried out in Ref.\cite{Salemi2022}, with the corresponding values from both theoretical works along with our experiment summarised in Table III.

\begin{table}[H]
    \centering
    \caption{Comparison of experimental and theoretical spin Hall conductivity values for Ni}
    \renewcommand{\arraystretch}{1.2}
    \setlength{\tabcolsep}{4pt}
    \footnotesize
    \begin{tabular}{l c c}
        \hline\hline
        \textbf{Reference} & $\sigma_{\mathrm{SHE}}^\perp$ $[\frac{\hbar}{e}]$ ($\Omega^{-1} \text{cm}^{-1}$) & $\sigma_{\mathrm{SHE}}^\parallel$ $[\frac{\hbar}{e}]$ ($\Omega^{-1} \text{cm}^{-1}$) \\
        \hline
        H-shaped & $3800 \pm 400$ & $2000 \pm 400$ \\
        L-shaped & $3000 \pm 400$ & $2000 \pm 300$ \\
        Amin et al. \cite{amin_intrinsic_2019} & $1688$ & $960$ \\
        Salemi et al. \cite{Salemi2022} & $1575$ & $824$ \\
        \hline\hline
    \end{tabular}
    \label{tab:SHC_comparison}
    \vspace{-1.1em}
\end{table}

Our spin Hall conductivities ($\sigma_\mathrm{SHE}=\theta_\mathrm{SHE}/\rho$), in both geometries are much larger than expected, by approximately a factor of 2 compared to the theory - although notably the ratio between them is comparable to the expected value. There are a number of possible explanations for this. Our values for $\lambda_\mathrm{s}$ are short in both cases. For the $\sigma_\mathrm{SHE}$ value extraction in Ni, we are using a $\lambda_\mathrm{s}$ in the Py injector of 2.6\,nm, which is around half the theoretically expected value. A larger value of $\lambda_\mathrm{s}$ in the injector would correspond to a lower $\sigma_\mathrm{SHE}$  in Ni, due to increased spin injection. Indeed, if we were to use the theoretical value of $\lambda_\mathrm{s}$ in the Py from \cite{Salemi2022} of 5.5\,nm, without changing the value of the Sharvin resistance we would halve our obtained $\sigma_\mathrm{SHE}$  in Ni. As well as this we have treated both interfaces (Cu/Py and Cu/Ni) as being identical, but this may not be the case. Furthermore, our model ignores additional interface effects, such as spin memory loss, which would have an effect on the result but are difficult to determine accurately in a specific device \cite{Liu2022-az}. Most importantly, we have seen that the precise value of the interface resistance is critical, but difficult to determine. We have used a single value from literature in our analysis, but given its large effect on $\lambda_\mathrm{s}$ and by extension the $\sigma_\mathrm{SHE}$, this is a key source of uncertainty. 

\section{Conclusion}
In conclusion, we have independently characterized the $\lambda_\mathrm{s}$ and $\theta_\mathrm{SHE}$ of Py and Ni as a function of the relative orientation of the fixed spin current polarization and the rotating magnetization. We perform these experiments in devices with transparent and Ohmic interfaces, and spin relaxation lengths, resistivities, and polarization of all materials measured in situ. 

We use two different device geometries, one based on previous experimental work (the H-shaped device) \cite{cosset-cheneau_electrical_2022}, which has the advantage of allowing for a larger signal due to close proximity of the injecting and absorbing element, as well as allowing the in-situ measurement of the magnetization dependent absorption of the spin current. We repeat the spin-charge interconversion measurement in a novel geometrical configuration (L-shaped device), which allows us to remove any possible artefacts due to non-collinear magnetization at the tips of the FMs, achieving very similar results. We furthermore see a high degree of reproducibility across multiple samples using both geometries. 

All the results are analysed using 3D FEM simulations, allowing us to account for the precise geometries used and giving us an accurate description of the behaviour of $\lambda_\mathrm{s}$ and $\theta_\mathrm{SHE}$ in Ni and Py, which are very similar between the two geometries used. We measure a large anisotropy in $\lambda_\mathrm{s}$ for both materials (Fig. \ref{fig:Results}a,b), although it is more anisotropic in Py. Similarly, we see a large anisotropy in $\theta_\mathrm{SHE}$ (Fig. \ref{fig:Results}c,d), which changes in the opposite direction to $\lambda_\mathrm{s}$, meaning that $R_\mathrm{ISHE}$ remains relatively constant, highlighting the importance of measuring both quantities independently. Finally, we compare our values of the spin Hall conductivity in Ni to those from two theoretical works. The ratio of $\sigma_\mathrm{SHE}^\mathrm{\parallel}$ to $\sigma_\mathrm{SHE}^\mathrm{\perp}$ is similar, although our absolute values are almost twice that of the calculated ones in Ni (Table III).
\begin{acknowledgments}
The authors acknowledge funding from MICIU/AEI/10.13039/501100011033 (Grant CEX2020-001038-M), from MICIU/AEI and ERDF/EU (Projects PID2021-122511OB-I00 and PID2021-128004NB-C21), from the European Union's Horizon 2020 research and innovation programme under the Marie Sk\l{}odowska-Curie grant agreement No. 955671. J.M. acknowledges funding by the Department of Education of the Basque Government under the Pre-doctoral Programme for the training of non-doctoral research staff.
\end{acknowledgments}

\section*{Data Availability}

The data that support the findings of this study are openly available~\cite{zenodo_data_2024}.

\setlength{\parskip}{1.3em}
\setlength{\parindent}{0pt}
\raggedbottom

\renewcommand{\thesection}{S\arabic{section}}
\renewcommand{\thefigure}{S\arabic{figure}}
\renewcommand{\figurename}{Fig.}
\setcounter{section}{0}
\setcounter{figure}{0}

\clearpage
\mbox{}
\thispagestyle{empty}
\onecolumngrid

\begin{center}
    \textbf{\LARGE Supplementary Information}
\end{center}

\renewcommand{\thefigure}{S\arabic{figure}}
\renewcommand{\figurename}{Fig.}
\renewcommand{\thesection}{S\arabic{section}}

\section{Material Parameters}

\begin{figure}[H]
    \centering
    \includegraphics[width=0.7\linewidth]{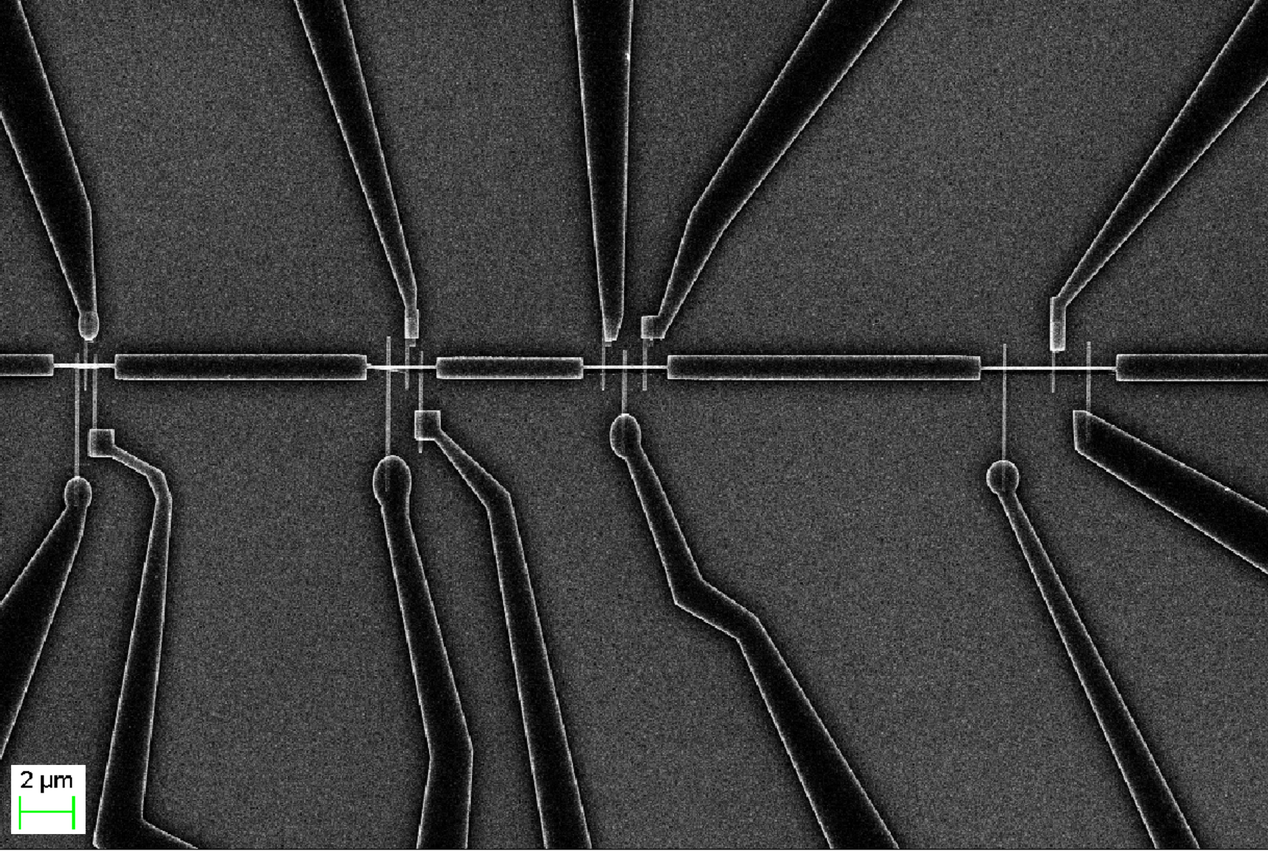}
    \caption{SEM image of a reference LSV device.}
    \label{fig:SI1}
\end{figure}

\begin{figure}[H]
    \centering
    \includegraphics[width=0.8\linewidth]{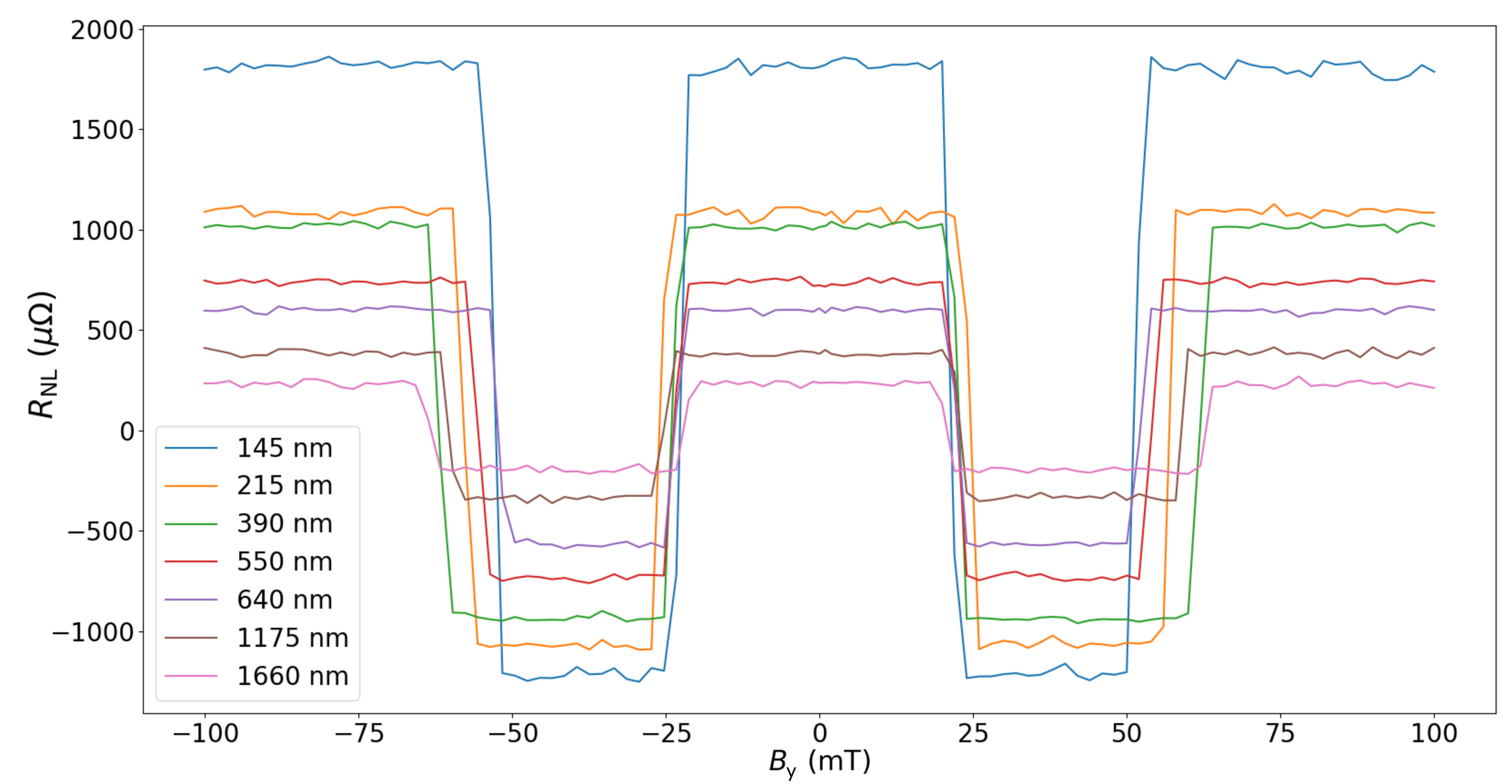}
    \caption{$R_\mathrm{NL}$ as a function of $B_\mathrm{x}$ at different $L$. The LSV with \(L\)\,=\,145\,nm has a non-zero baseline signal due to very close proximity and thus charge current from the injector arriving at the detector. All measurements are performed at 10\,K.}
    \label{fig:SI2}
\end{figure}

We measure the non-local resistance, $R_\mathrm{NL}$, in a series of LSVs with different distances, $L$, between the FM injector and detector. The device used for these measurements is shown in Fig. \ref{fig:SI1}. This reference device is fabricated on the same chip as the devices used in the paper, ensuring that the Cu channel and Py spin injectors have the same dimensions and properties. \(R_{\text{NL}}\) for each LSV, with different $L$, is shown in Fig.~\ref{fig:SI2}. The spin signal, $\Delta R_\mathrm{NL}$, is defined as the difference between $R_\mathrm{NL}$ in the parallel and anti-parallel configuration of the FM electrodes.

\begin{figure}[H]
    \centering
    \includegraphics[width=0.5\linewidth]{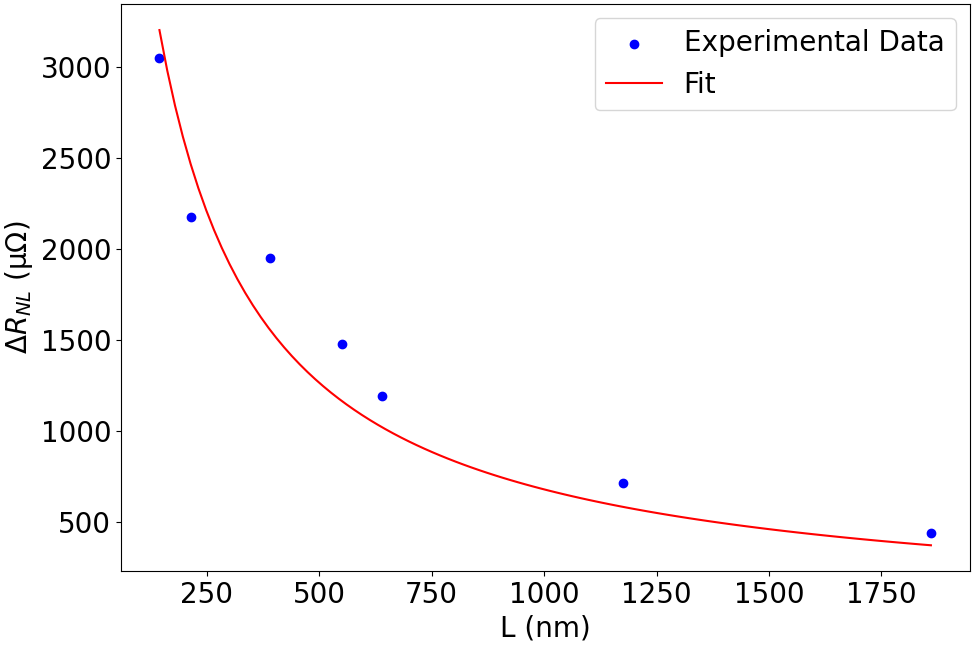}
    \caption{Fitting of the spin signal, $\Delta R_\mathrm{NL}$,  as a function of the distance, \(L\), between the injector and detector to Eq.\ref{eq:RNL}}
    \label{fig:SI3}
\end{figure}

The value of $\Delta R_\mathrm{NL}$ decays rapidly with distance, as shown in Fig. \ref{fig:SI3}. At the smallest distance, a non-zero baseline signal is present due to stray charge current reaching the detector. This baseline is subtracted for all fitting. 

Approximating the device as 1D, the length dependence of $\Delta R_\mathrm{NL}$ can be modelled by the equation\cite{sagasta_spin_2017}:

\begin{equation}
\label{eq:RNL}
\Delta R_{\text{NL}}(L, \alpha_{\text{Py}}, \lambda_\mathrm{s}^{\text{Cu}}, \lambda_\mathrm{s}^{\text{Py}}) 
= \frac{
  4\,\alpha_{\text{Py}}^2\,R_\mathrm{s}^\mathrm{Cu}
}{
  \bigl(2 + \tfrac{R_\mathrm{s}^\mathrm{Cu}}{R_\mathrm{s}^\mathrm{Py}}\bigr)^2 \, e^{L / \lambda_\mathrm{s}^{\text{Cu}}}
  \;-\;
  \bigl(\tfrac{R_\mathrm{s}^\mathrm{Cu}}{R_\mathrm{s}^\mathrm{Py}}\bigr)^2\, e^{-L / \lambda_\mathrm{s}^{\text{Cu}}}
}
\,,
\end{equation}
where \(R_\mathrm{s}^\mathrm{Cu} = \tfrac{\lambda_\mathrm{s}^{\text{Cu}} \rho_{\text{Cu}}}{w_{\text{Cu}} \, t_{\text{Cu}}}\) 
and \(R_\mathrm{s}^\mathrm{Py} = \tfrac{\lambda_\mathrm{s}^{\text{Py}} \rho_{\text{Py}}}{w_{\text{Cu}} \, w_{\text{Py}} \,(1 - \alpha_{\text{Py}}^2)}\) are the spin resistances in each material. Here, \(w_{\text{Cu}}\) and \(w_{\text{Py}}\) denote the widths of the Cu and Py layers, respectively, while \(t_{\text{Cu}}\) represents the thickness of the Cu channel. $\alpha_\mathrm{Py}$ is the polarisation of Py, and $\lambda_\mathrm{s}^\mathrm{Cu/Py}$ are the respective spin diffusion lengths. A fitting with this equation is shown in Fig. \ref{fig:SI3}.

\begin{figure}[H]
    \centering
    \includegraphics[width=0.9\linewidth]{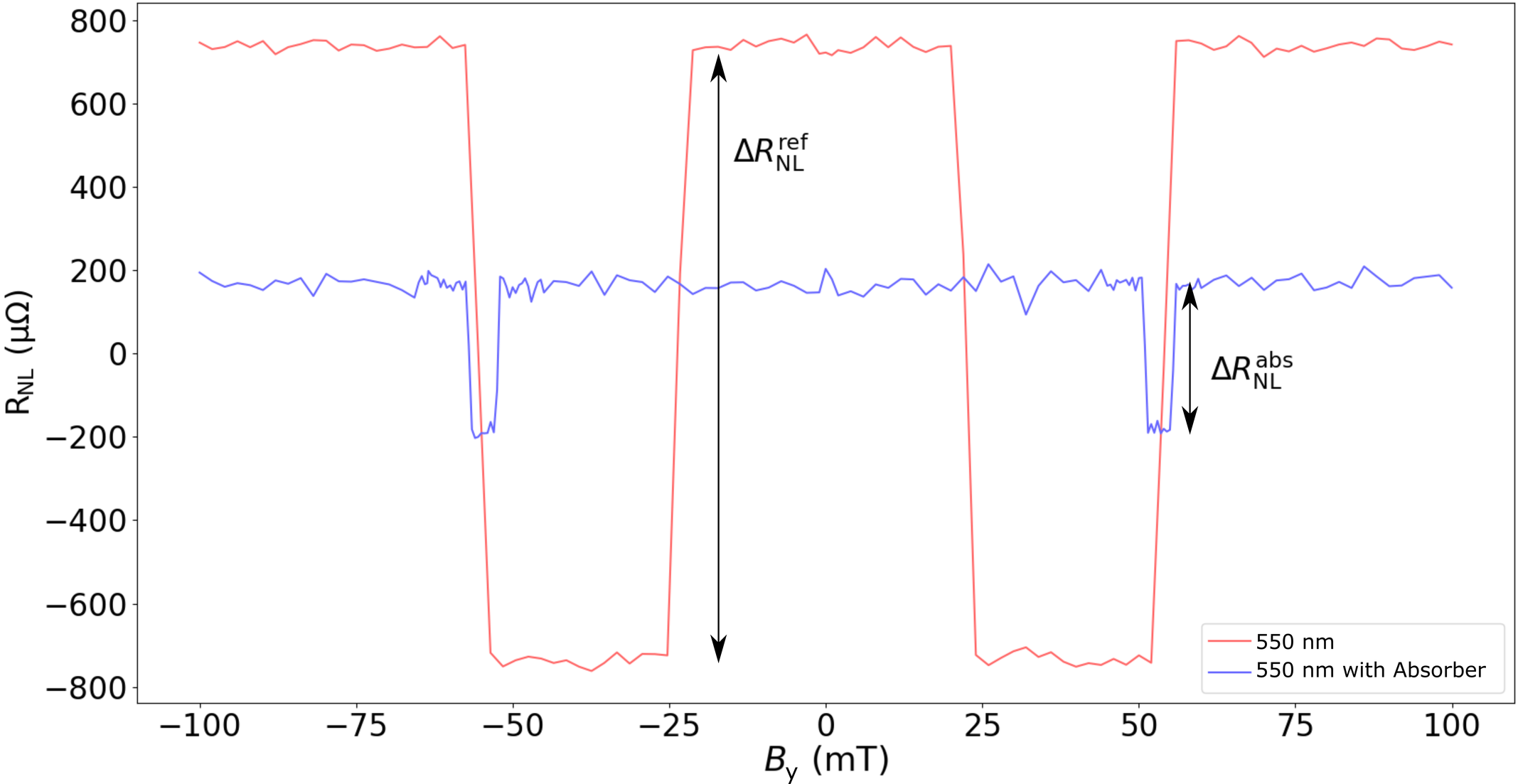}
    \caption{Comparison of LSV signals at 10\,K, with the same $L$ (550\,nm), with and without an absorber.}
    \label{fig:SI4}
\end{figure}

However, the effects of \(\alpha_{\text{Py}}\) and \(\lambda_\mathrm{s}^{\text{Py}}\) on the signal are similar, leading to a non-unique solution. To constrain the problem, we incorporate data from Fig. \ref{fig:SI4}, where we measure the spin signal ratio (\(\eta\)) for an LSV with (\(\Delta R_{\mathrm{NL}}^{\mathrm{abs}}\)) and without (\(\Delta R_{\mathrm{NL}}^{\mathrm{ref}}\)) an absorber. The ratio of the two spin signals (\(\eta = \frac{\Delta R_{\mathrm{NL}}^{\mathrm{abs}}}{\Delta R_{\mathrm{NL}}^{\mathrm{ref}}}\)) in Fig. \ref{fig:SI4} is given by:

\begin{equation}
\label{eq:eta}
\eta(L, \lambda_\mathrm{s}^{\text{Cu}}, Q_M, Q_{\text{Py}}) 
= \frac{
  2\,Q_M \,
  \bigl[
    \sinh\bigl(\tfrac{L}{\lambda_\mathrm{s}^{\text{Cu}}}\bigr) 
    + 2\,Q_{\text{Py}} \, e^{L / \lambda_\mathrm{s}^{\text{Cu}}}
    + 2\,Q_{\text{Py}}^2\, e^{L / \lambda_\mathrm{s}^{\text{Cu}}}
  \bigr]
}{
  \cosh\bigl(\tfrac{L}{\lambda_\mathrm{s}^{\text{Cu}}}\bigr) 
  - 1
  + 2\,Q_M\,\sinh\bigl(\tfrac{L}{\lambda_\mathrm{s}^{\text{Cu}}}\bigr)
  + 2\,Q_{\text{Py}} \bigl[
    e^{L / \lambda_\mathrm{s}^{\text{Cu}}}\,(1 + Q_{\text{Py}})\,(1 + 2\,Q_M)
    - 1
  \bigr]
}
\,,
\end{equation}
where \(Q_{\text{Py}} = \tfrac{R_{\text{Py}}}{R_{\text{Cu}}}\) and \(Q_M = \tfrac{R_{\text{M}}}{R_{\text{Cu}}}\). Here, $R_\mathrm{M}$ is the spin resistance of the absorber, $R_{\text{M}} = \tfrac{\lambda_\mathrm{s}^{\text{M}} \rho_{\text{M}}}{w_{\text{Cu}}\, w_{\text{M}}\, (1 - \alpha_{\text{M}}^2)}$

By simultaneously fitting the length dependence (Fig. \ref{fig:SI3}) and incorporating this ratio, a unique solution is achieved, following the method in~\cite{sagasta_spin_2017}.

The absorber here is Py, so \(Q_{\text{Py}} = Q_M\). By fitting the length dependence using this analytical model (Eq. 1) together with the absorber/no-absorber constraint in the case of the Py absorber (Eq. 2), we obtain a unique solution: \(\alpha_{\mathrm{Py}} = 0.28 \pm 0.01\), \(\lambda_\mathrm{s}^{\mathrm{Py}} = 4.5 \pm 0.2\,\mathrm{nm}\), and \(\lambda_\mathrm{s}^{\mathrm{Cu}} = 1300 \pm 100\,\mathrm{nm}\). 

\section{3D Simulations}
The 1D approach is suitable for the conventional LSV geometry of the reference device but becomes less accurate for the SHE measurement devices, which feature more complex geometries. We therefore also perform 3D finite-element simulations (3D-FEM), similar to the method in~\cite{zahnd_comparison_2016}, for both the reference device and the absorber configuration. This 3D simulation method also allows us to incorporate the Sharvin resistance at the Cu/Py and Cu/Ni interfaces, as discussed in the main text. 

\begin{figure}[htpb]
    \centering
    \includegraphics[width=1\linewidth]{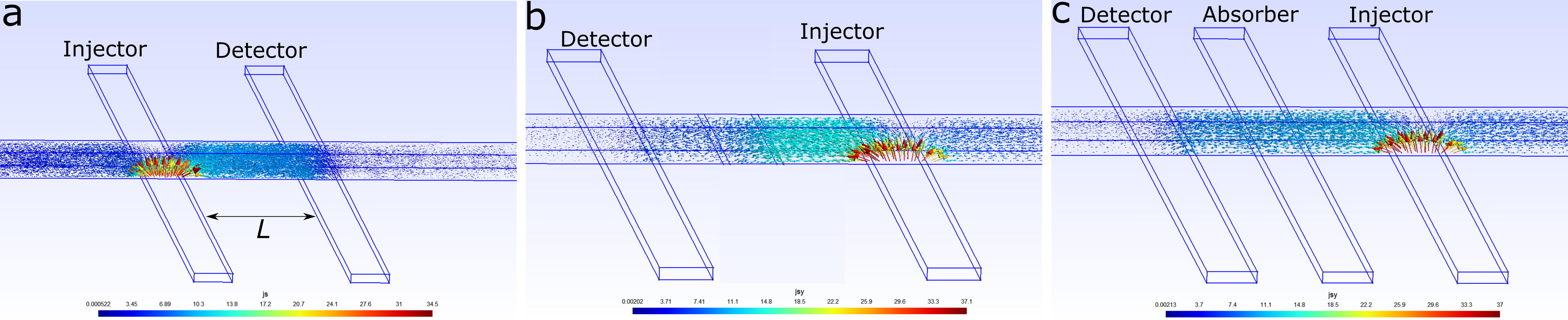}
\caption{Summary of the geometries used to extract spin diffusion parameters, $\lambda_\mathrm{s}^\mathrm{Cu}$, $\lambda_\mathrm{s}^\mathrm{Py}$ and $\alpha^\mathrm{Py}$ consisting of an $L$ dependant LSV (a), an LSV with fixed length 
 (b), and an LSV with an identical fixed length and an absorbing Py element between the injector and detector ($L$). Colour scale shows the spin current in each case.}
    \label{fig:Length_Abs_Nabs}
\end{figure}

\begin{figure}[H]
    \centering
    \includegraphics[width=1\linewidth]{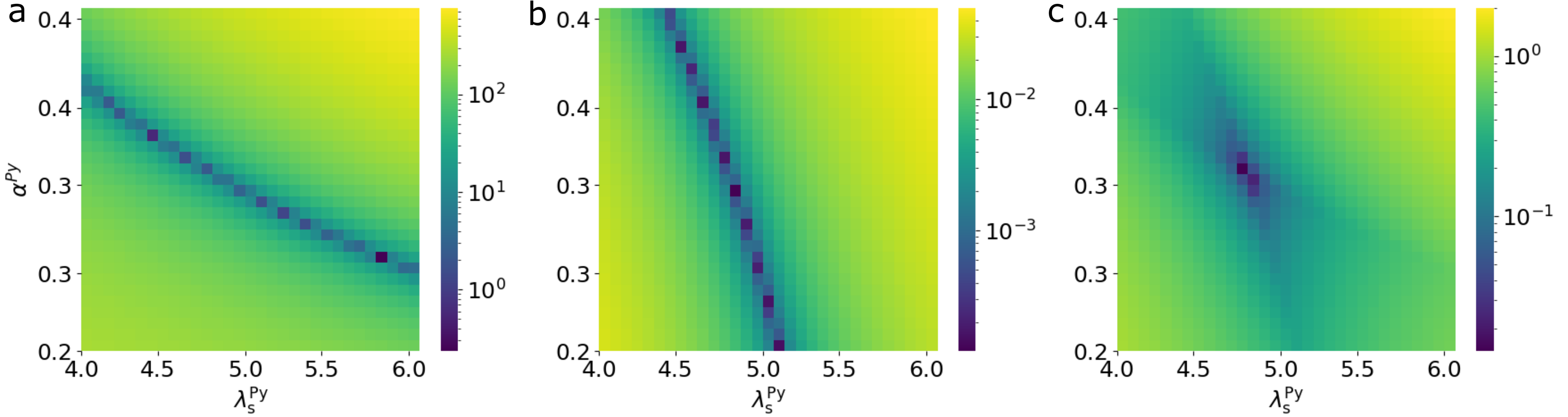}
    \caption{a) RMSE between the experimental length dependence measurement and the simulated one over a range of ($\lambda_\mathrm{s}^\mathrm{Py}$ and $\alpha^\mathrm{Py}$) for a fixed $\lambda_\mathrm{s}^\mathrm{Cu}$ of 1350\,nm. b) The ratio, $\eta$, between the LSV with and without an absorber, showing the deviation from the experimental values. c) Addition of the data in (a) and (b), with the intersection showing the optimized values for $\lambda_\mathrm{s}^\mathrm{Py}$ and $\alpha^\mathrm{Py}$ where all spins are parallel to magnetisations.}
    \label{Length_Abs_Heatmap}
\end{figure}

First, we simulate the length dependence of $R_\mathrm{NL}$, using a range of experimental $L$ values. We choose five evenly spaced $L$ values in the range of experimental measurements from Fig. \ref{fig:SI3}, where we use the 1D fitted curve as an interpolation of the data, letting us choose a corresponding $\Delta R_\mathrm{NL}$ for each value of length. Then we run the simulation shown in Fig. \ref{fig:Length_Abs_Nabs}a for each value of $L$, as well as an array of values for \(\lambda_\mathrm{s}^{\text{Cu}}\), \(\lambda_\mathrm{s}^{\text{Py}}\), and \(\alpha_{\text{Py}}\). We want the output to match as closely as possible to the experimental curve, so we take the simulated $\Delta R_\mathrm{NL}({L})$ curve, and compare it to the experimental curve, calculating the root mean square error (RMSE) between them. A plot of this error, for a fixed value of $\lambda_\mathrm{s}^\mathrm{Cu}$, is shown in Fig. \ref{Length_Abs_Heatmap}a.

As is clearly visible, there is a minimum corresponding to a close match between the experimental and simulated data, however it is not unique - a family of solutions exist along a curve, meaning just from this data we cannot extract unique values of $\lambda_\mathrm{s}^\mathrm{Py}$ and $\alpha_\mathrm{Py}$. This is the same problem faced in the 1D analytical approach. We therefore take advantage of the definition of $\eta$ in Eq. 2, and also simulate the situation where we have a LSV with and without an absorbing element (Fig. \ref{fig:Length_Abs_Nabs}b,c) in between the injector and detector corresponding to the data in Fig. \ref{fig:SI4}. Again we perform simulations of the two systems over a range of \(\lambda_\mathrm{s}^{\text{Cu}}\), \(\lambda_\mathrm{s}^{\text{Py}}\), and \(\alpha_{\text{Py}}\), and extract the corresponding range of values for $\Delta R_{\mathrm{NL}}^{\mathrm{abs}}$ and $\Delta R_{\mathrm{NL}}^{\mathrm{ref}}$. We divide these values to get the simulated range of $\eta$. We compare this to the experimental value of $\eta$ from Fig. \ref{fig:SI4}, and plot the absolute difference in Fig. \ref{Length_Abs_Heatmap}b. Again, this gives a family of possible values for $\lambda_\mathrm{s}^\mathrm{Py}$ and $\alpha_\mathrm{Py}$, so by overlapping the two methods we can find a unique solution for $\lambda_\mathrm{s}^\mathrm{Py}$, $\alpha_\mathrm{Py}$ and $\lambda_\mathrm{s}^\mathrm{Cu}$. This method is carried out for an array of $\lambda_\mathrm{s}^\mathrm{Cu}$, with the lowest global deviation from the experimental data being shown in Fig. \ref{Length_Abs_Heatmap}, corresponding to $\lambda_\mathrm{s}^\mathrm{Cu} = 1350\, \mathrm{nm}$. By simulating the device over a wide range of \(\lambda_\mathrm{s}^{\text{Cu}}\), \(\lambda_\mathrm{s}^{\text{Py}}\), and \(\alpha_{\text{Py}}\), and minimizing the difference between experiment and simulation, we find:
\[
\lambda_\mathrm{s}^{\text{Cu}} = 1350 \pm 50\,\mathrm{nm}, \quad
\lambda_\mathrm{s}^{\text{Py}} = 4.8 \pm 0.2\,\mathrm{nm}, \quad
\alpha_{\text{Py}} = 0.33 \pm 0.02.
\]

The small deviation from the 1D model (\(\lambda_\mathrm{s}^{\mathrm{Cu}} = 1300 \pm 100\,\mathrm{nm}\), \(\lambda_\mathrm{s}^{\mathrm{Py}} = 4.5 \pm 0.2\,\mathrm{nm}\), \(\alpha_{\mathrm{Py}} = 0.28 \pm 0.01\)), is expected since the system is not perfectly 1D, with contact widths of 130\,nm, there is a gradient of spin potential across the interfaces which is not accounted for in the 1D model, as well as a potential gradient through the thickness of the Cu channel (90\,nm). Nevertheless, the two approaches agree quite well. For the case of the Ni absorber, we use the H-shaped device from the main text. Since we already know that $\alpha_\mathrm{Ni}=0$, we have just $\lambda_\mathrm{Ni}$ as a variable. We can use the previously calculated values of $\lambda_\mathrm{s}^{\mathrm{Py}}$ and $\alpha_\mathrm{Py}$ for the injector and detector elements, and then vary the value of $\lambda_\mathrm{s}^{\mathrm{Ni}}$ to match the experimental $R_\mathrm{NL}$. In this case we will get different values depending on the magnetisation of the nickel, as shown in Table II of the main text. 

All of these simulations are done with a perfectly transparent interface between all $ \text{Py} $ and $ \text{Cu} $ elements, but this is not physically accurate due to the Sharvin conductance providing a maximum conductance value of the interface of $ 1.2 \times 10^{15} \, \text{f} \Omega^{-1} \text{m}^{-2} $.\cite{cosset-cheneau_measurement_2021} The 3D simulation result corresponds to a value of the spin conductance, $ G_{\text{s}} $, in the $ \text{Py} $ since we are ignoring the interface. This $ G_{\text{s}} $ is defined as:

\begin{equation}
\label{eq:Geff}
G_{\text{s}} = \frac{1}{R_{\text{s}}} = \frac{(1 - \alpha_{\text{Py}}^2)}{\rho_{\text{Py}} \lambda_{\text{s}}^{\text{Py}}}
\end{equation}

\begin{figure}[htpb]
    \centering
    \includegraphics[width=0.85\linewidth]{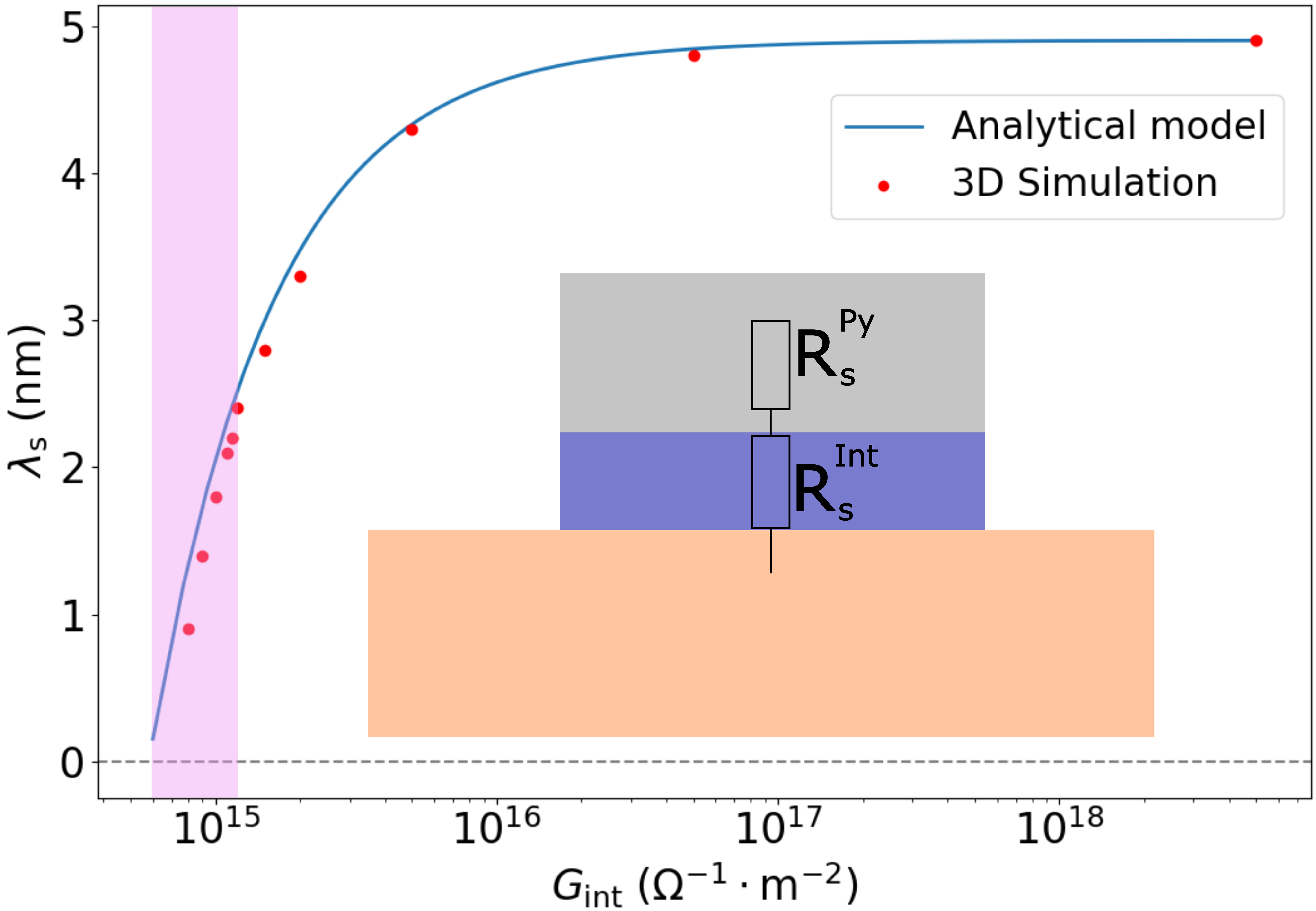}
\caption{For a range of interface conductance ($G_\mathrm{int}$) values, we calculate the value of  $\lambda_\mathrm{s}$ needed for the total spin resistance to match that calculated by 3D FEM simulation. This is done based on the series resistor model shown in the inset. The highlighted region corresponds to the range of possible $G_\mathrm{int}$ values based on a lower limit set by our experimentally measured interface conductivity and an upper limit from the theoretical Sharvin conductivity.}
    \label{fig:Lambda_Interface_3D_1D}
\end{figure}

However, in reality, due to the Sharvin resistance at the interface ($R_\mathrm{Sh}$), there is an additional resistance felt by the spins as they enter the $\text{FM}$. This can be modeled as a series resistor model (Fig.~\ref{fig:Lambda_Interface_3D_1D}), with the first resistance caused by the interface ($R_{\text{Sh}}$), and the second due to the $\text{Py}$ ($R_{\text{s}}$). The effective spin conductance is therefore defined by:

\begin{equation}
\label{eq:Geff_series}
\frac{1}{G_{\text{eff}}} = \frac{1}{G_{\text{Sh}}} + \frac{1}{G_{\text{s}}}
\end{equation}

Using the calculated value of $G_\mathrm{s}$ from the 3D FEM fitting of the reference device (Fig. \ref{Length_Abs_Heatmap}), with its corresponding values of $\lambda_\mathrm{s}$ and $\alpha_\mathrm{s}^\mathrm{Py}$ (Eq.\ref{eq:Geff}), we can analytically calculate the $\lambda_\mathrm{s}$ in the Py, which gives the correct $G_\mathrm{eff}$ when considered in series with the interface. Alternatively, we can set the value of $G_\mathrm{Sh}$, and run the simulation with this value to extract a corresponding $\lambda_\mathrm{s}$ which matches the experimentally measured $R_\mathrm{NL}$. Both methods agree closely, and are summarised in Fig. \ref{fig:Lambda_Interface_3D_1D}.

In general, the interface resistance is very important to quantifying the behaviour of such systems, but is very difficult to measure accurately. We do measure the interface resistance in a device, as described in Note S3, but this can only give us an upper bound to the interface resistance. A Sharvin conductivity of $1.2 \times 10^{15} \, \text{f} \Omega^{-1} \text{m}^{-2}$ corresponds to an overall interface resistance of 0.05 $\Omega$ in our junction. This provides a lower limit on the interface resistance. However, from the experimental measurement, we extract an upper limit of 0.1 $\Omega$, corresponding to an interface conductance ($G_\mathrm{int}$) of $0.6 \times 10^{15} \, \text{f} \Omega^{-1} \text{m}^{-2}$. This range is marked on Fig. \ref{fig:Lambda_Interface_3D_1D}, and corresponds to a significant range of $\lambda_\mathrm{s}$ values.

\begin{figure}[htpb]
    \centering
    \includegraphics[width=1\linewidth]{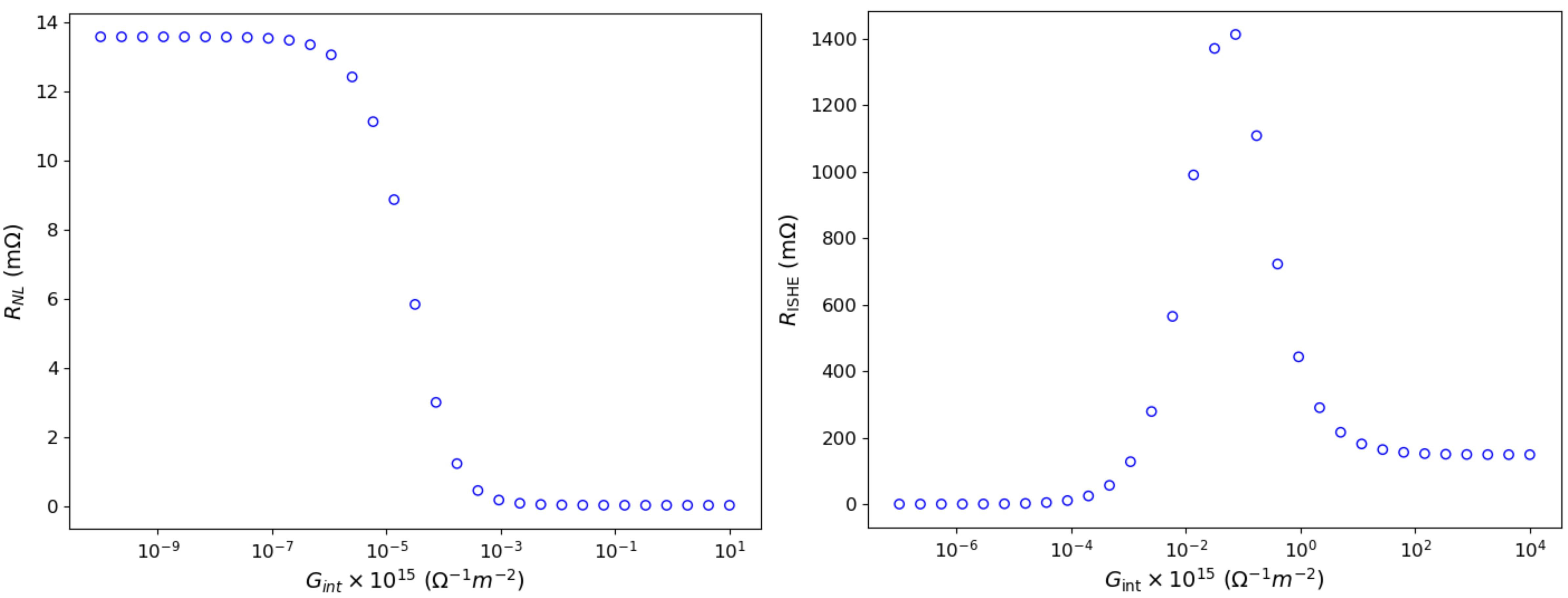}
\caption{a) The $R_\mathrm{NL}$ simulated in the LSV configuration of the H-shaped device as a function of interface conductance, where all interface conductances are equal. At large values of $G_\mathrm{int}$, we converge on the transparent limit of $\approx 0.1 \mathrm{m}\Omega$ b) $R_\mathrm{ISHE}$ again measured in the H-shaped device, as a function of interface conductance. All simulations are carried out a a single large value of $B_\mathrm{x}$, and correspond to approximately $\Delta R_\mathrm{NL}/2$, although the exact values are arbitrary and meant just to illustrate the importance of choosing an appropriate $G_\mathrm{int}$.}
    \label{fig:RNL_RISHE_Gint}
\end{figure}

To further highlight the importance of considering the interface resistance, we plot, for the H-shaped device, the values of $R_\mathrm{NL}$ in the LSV configuration, as well as $R_\mathrm{ISHE}$, both of which show order of magnitude differences in the non-local resistance depending on the interface resistance used, but with both dependences being different (see Fig. S8).

Given that we cannot accurately gauge the value of the interface resistance, when conducting the simulations on the H-shaped device to extract $\lambda_\mathrm{s}^\mathrm{\perp}$, we use the value of $\lambda_\mathrm{s}^\mathrm{\parallel}$ (2.6 nm) corresponding to the literature value of Sharvin resistance of $1.2 \times 10^{15} \, \text{f} \Omega^{-1} \text{m}^{-2}$. This value corresponds to the injector and detector in the H-shaped device, as well as to the high field value for the absorber (when all elements are aligned to the field). This introduces quite a large error on the absolute value of $\theta_\mathrm{SHE}$, however the ratio between $\theta_\mathrm{SHE}^\mathrm{\parallel}$ and $\theta_\mathrm{SHE}^\mathrm{\perp}$ will be affected much less, and this is the principal quantity of interest to quantify the SHE anisotropy.

\FloatBarrier

\section{Interface Resistance}

\begin{figure}[H]
    \centering
    \includegraphics[width=1\linewidth]{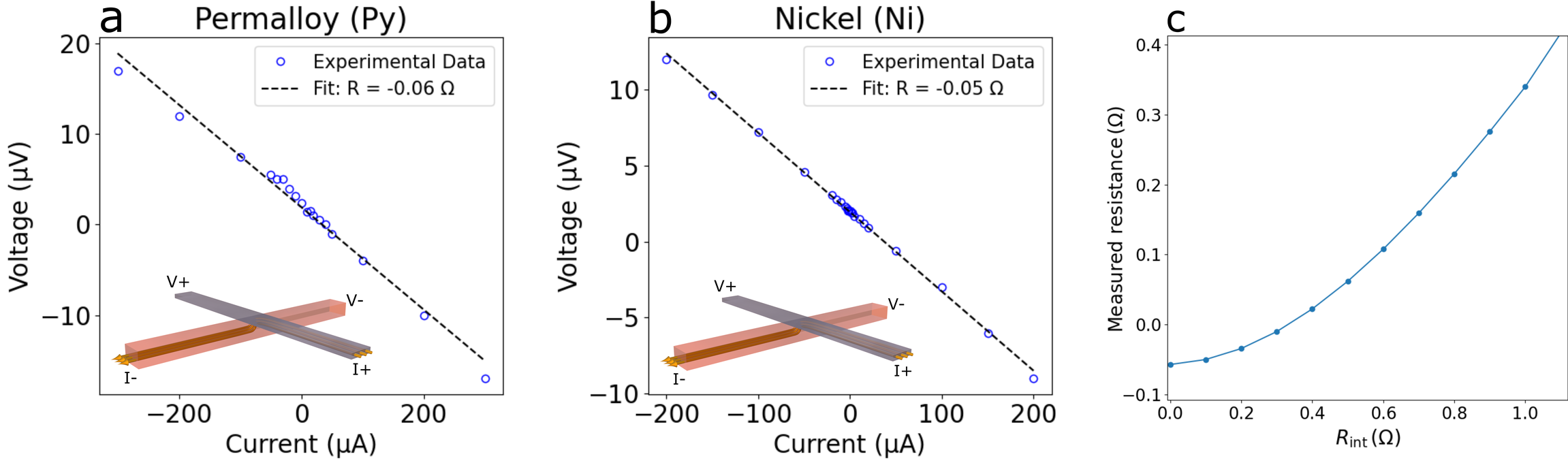}
    \caption{Experimental interface resistance of (a) Cu/Py and (b) Cu/Ni. The inset shows the reference device geometry and the measurement configuration. (c) Data from 3D FEM simulation of the Cu/Py system. Here, we simulate the measured resistance output as $R_\mathrm{int}$ is varied, which perfectly captures the negative values observed for highly transparent interfaces. Similar results are found for the Cu/Ni system}.
    \label{fig:SI6}
\end{figure}

We extract the interface resistance ($R_\mathrm{int}$) of our devices to ensure a transparent interface between the Cu and the FM. The measurement configuration is shown in Fig.~\ref{fig:SI6}a and b (insets), with a Cu/FM cross, where the FM is Py or Ni. This is done over a range of applied currents (to ensure the interface is Ohmic), and the data are fitted so that the slope corresponds to the measured interface resistance via Ohm's law. This slope is found to be negative, which is expected from highly transparent interfaces. To confirm this, we perform a 3D simulation of the device in Fig.~\ref{fig:SI6}c, plotting the measured resistance vs. $R_\mathrm{int}$. We see that, at very low $R_\mathrm{int}$ we return a small negative value of the measured resistance due to the current distribution across the interface. This justifies treating the interface as transparent in all of our analyses, although we can only give an upper bound for $R_\mathrm{int}$ by this method, with $R_\mathrm{Sh}$ being the lower limit.

\section{Field Dependence of Magnetisation}

\begin{figure}[H]
    \centering
    \includegraphics[width=1\linewidth]{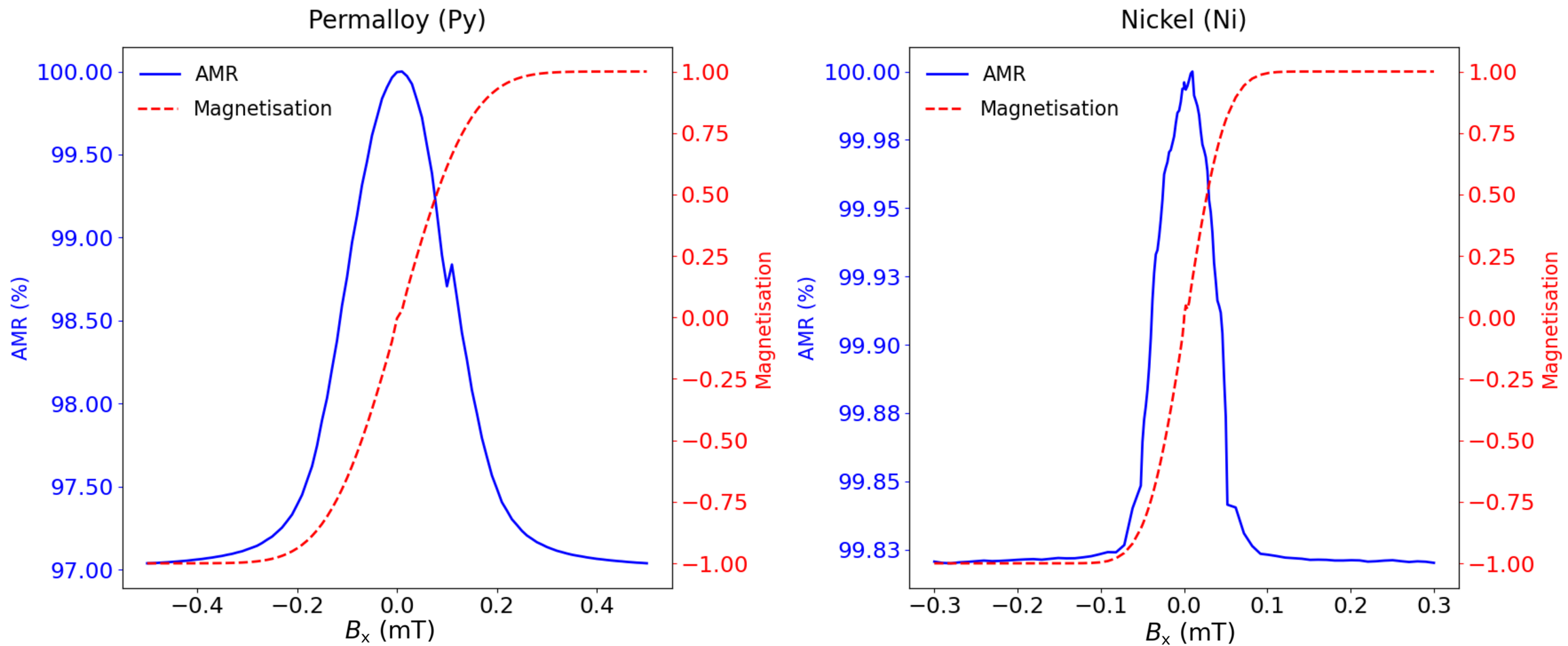}
    \caption{AMR measurement of Py and Ni (blue), showing the corresponding hard-axis magnetisation (red).}
    \label{fig:SI8}
\end{figure}
For the analysis of the magnetisation dependence of \(\theta_{\text{SHE}}\) and \(\lambda_\mathrm{s}\), we need the behaviour of the magnetisation with applied field, as this is the directly controlled quantity. A two-point measurement of the FM absorber in the H-shaped devices was carried out with the field along the hard axis. This yields a standard anisotropic magnetoresistance (AMR) curve for the two materials; by fitting it to a Lorentzian function and multiplying by the field direction, we obtain a good analytical approximation for the magnetisation behaviour, which is then used in subsequent simulations. Note that Ni is much softer than Py, and is more prone to non-ideal behaviour and hysteresis at low field. This accounts for both the shape of the Ni AMR curve and likely contributes to the fit in Fig.~2 of the main text being less accurate for Ni than for Py. Another possibility for the Ni case is that, due to the local environment at the Cu/Ni interface, the local magnetisation requires a larger field to orient along the hard axis, compared to the entire Ni structure on which AMR is measured. This would explain the deviation of the fitting for \(\lambda_\mathrm{s}\) in Ni at intermediate fields (Fig. 2c) of the main text.

\section{Spin accumulation in Py}
The background signal in the Py SHE measurements (Fig. 3a, main text), which is symmetric with magnetic field, is attributed to spin accumulation at the interface. In an ideal system, this should not lead to any signal in the SHE configuration (green circuit, Fig. 1a). However, the spin accumulation signal is significantly larger than the SHE signal, and any anisotropy at the Py/Cu interface can lead to a spin accumulation gradient which manifests as a non-local voltage. Even a small part of this accumulation being detected can therefore be a large effect on the scale of the SHE measurement as we see here. Other origins for this background cannot be completely ruled out, but in Fig. S11 we show a direct measurement of the spin accumulation at the Cu/Py interface in the H-shaped device, which matches the symmetric background in the SHE measurement (Fig. 3a, main text) well. This, coupled with the absence of such a background in Ni, is good evidence that the background is due to spin accumulation.

\begin{figure}[H]
    \centering
    \includegraphics[width=0.45\linewidth]{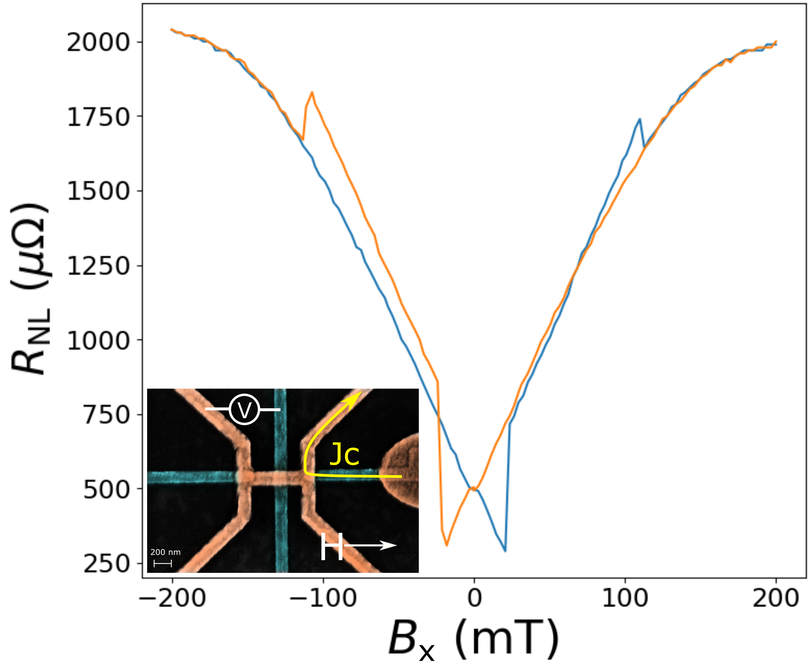}
    \caption{Spin accumulation at the Cu/Py absorber interface in the H-shaped device. This is equivalent to a standard LSV measurement with perpendicular injector and detector.}
    \label{fig:SI8a}
\end{figure}

\end{document}